\documentstyle[12pt]{article}
\begin{document}
\title{\rightline{\rm\normalsize{JHU-TIPAC-930013}}\vspace*{1.0cm}
Semiclassical Transition in $\phi^{4}$ Theory\thanks{Work
supported in part by the National Science Foundation
 under grant PHY-90-96198.}}
\author{Alexander Kyatkin\thanks{kyatkin@dirac.pha.jhu.edu}\\
Department of Physics and Astronomy\\
The Johns Hopkins University\\Baltimore MD 21218\vspace*{1.5cm}}
\date{April 1993}
\maketitle

\begin{abstract}
We have shown an example of semiclassical transition in
$\phi^{4}$ theory with positive coupling constant. This process
 can be described by the classical $O(4)$-invariant solution,
considered on a contour in the complex time plane. The transition
 is technically analogous to the one-instanton transition in the
 electroweak model. It is suppressed by the factor $\exp(-2S_{0})$
, where $S_{0}$ is  Lipatov instanton action. This process describes
 a semiclassical transition between two coherent states with much
  smaller number of particles in the initial state than in the final
 state. Therefore, it could be relevant to the problem of calculation
 of amplitudes for multiparticle production in $\phi^4$-type models.
\end{abstract}
\newpage

{\section {Introduction}}
\vspace{1cm}

Recently, considerable efforts have been made to calculate
amplitudes for multiparticle production in  weakly coupled
field theories . The study of this problem
 was initiated by the observation of the fact \cite {Ringwald}
 that baryon-number violating processes in the electroweak
 theory, associated with multiparticle production, could become
 relevant at energy scale $E\sim 10\,TeV$. This problem gave
 impulse to
 study multiparticle amplitudes in the simpler case of $\phi^4$
 model \cite {Cornwall,Shifman}
, considered before in the context of large orders
 of perturbation theory \cite {Lipatov}.

The semiclassical methods for computing such amplitudes in
 the electroweak theory use Euclidean classical solutions
of the equations of motion --  instantons \cite {tHooft}.
The similar calculations in $\phi^4$ theory \cite {Shifman}
are based on the existence of the instanton-like
solutions in $\phi^4$ model with negative coupling constant \cite
{Lipatov}.
 These instanton-like calculations show an exponential growth of the
total cross sections with energy in the leading order
 of perturbation theory around the instanton \cite {three,mod}. A
 naive extrapolation of these results
 to the high energy scale violates the unitarity bound for
 the cross section.

The problem is  that instanton solutions describe transitions
between vacuum states. The real process, however, describes
 the production of many final-state particles in high energy,
two-particle collisions. Obviously, we cannot ignore the
effects of the  external particles.
  Indeed, at high energies (for
 example energies in the order of the sphaleron
energy in the electroweak model \cite {Manton})
instanton calculations become inappropriate, which can be seen in
the fact
that corrections to the leading-order transition probability become
 large \cite{Rubakov1,mod}. Therefore, in order to estimate accurately
 the external particle effects we have to modify the instanton-based
approach.

Formally, we cannot calculate
 the transition probability for the process $two\rightarrow many$
particles in the
semiclassical manner at all, because of the
 non-semiclassical nature of the
initial two-particle state. Instead, as proposed in Ref.\cite
{Approach}, we can calculate the probability of
transition between a semiclassical initial
 state with a ``small'' number of
particles and a final semiclassical state with a ``large''
 number of particles.
The probability of such a transition can be considered
as some approximation to the two particle cross section
 in one-instanton sector and gives us an
 upper bound for this cross section.

In this approach the problem can be reduced to the
 solution of the classical
field equations with some specific boundary conditions,
determined by the initial and final states.
 Such a formalism, based on the coherent
state representation of the
 $S$-matrix elements \cite {Berezin},
has been used to find the transition probability for processes
mediated by so-called ``periodic instantons'' \cite {Rubakov3}.

Finding the  exact form of the relevant high-energy instanton-like
configuration is, however, a very difficult problem, even in the
massless limit of the theory. To avoid this problem, the formalism
 of Refs.\cite {Approach,Rubakov3} has been
recently modified \cite {Rubakov4}
 to use exact Minkowskian classical solutions, which can be easily
 found for a number of models. This modification allows one to calculate,
 in principle, the semiclassical scattering above
the sphaleron energy in the electroweak model.
 It has been shown in \cite {Rubakov4}, that a
 Minkowskian solution of the  $O(3)$-invariant two-dimensional
$\sigma$-model, analytically continued to the complex time plane, can
be used to
describe instanton-like processes in this model with a
``strong'' violation of the
number of particles ($n_{initial}<<n_{final}$).

In this paper we consider  four-dimensional
massless $\phi^4$ theory. While $\phi^4$ theory with positive coupling
constant does not allow direct instanton-like calculations (we have to use
 Lipatov's trick \cite {Lipatov} and consider first the
theory with negative coupling constant), the formalism of Ref. \cite
{Rubakov4} can be used for direct semiclassical calculations
in this theory.

We show that $\phi^4$ theory allows
 a semiclassical transition even for the case of
positive coupling constant. This transition is described by
 a classical $O(4)$-invariant solution, considered on a contour
 in the complex time plane. The ``type'' of the transition is
 determined by the position of this contour with respect to the positions
 of the singularities of the classical solution.

 We consider the transition technically
 analogous to the one-instanton
 transitions in the electroweak model. It
is suppressed by the factor $\exp {(-2S_{0})}$, where $S_{0}$
is equal to Lipatov  instanton action -- the action of
the classical solution in the Euclidean theory with  negative
coupling constant \cite {Lipatov}. To interpret the process,
we analyze  a similar ``transition'' in one-dimensional
quantum mechanics.

This process describes a classically-forbidden
 transition between two coherent states with a much smaller number
of particles in the initial state than in the final state
 -- $n_{final}\sim n^{5/7}_{initial}/\lambda^{2/7}$
(where $\lambda$ is a small coupling constant). Therefore, it
could be relevant to the calculation of amplitudes for multiparticle
 production in $\phi^4$-type models.
We suppose that the contribution of such a process must be included
 into the corresponding multiparticle amplitude and, probably, can
 slow down the factorial growth of the
perturbative amplitude \cite {Cornwall}.

The paper is organized as follows.
 In the second section we describe the basic formalism \cite {Rubakov4}.
The third section is devoted to the description of the classical
$O(4)$-invariant
solutions of the $\phi^4$ theory and calculation of the transition probability.
As an example, we consider also the theory with  negative coupling
constant, where the results can be compared with
previous calculations \cite {Tinyakov}.
In section (4) we find the initial and final coherent states as
 asymptotics of the classical solution and the corresponding numbers of
the initial and final particles.
The last section contains concluding remarks.

\vspace{1cm}

{\section {Formalism}}

\vspace{1cm}

 In this section we describe the slightly modified
formalism  of the Refs. \cite {Rubakov3,Rubakov4},
based on the coherent state representation  of the $S$-matrix
elements \cite {Berezin}.
The $S$-matrix element in this representation
 is a generating functional for transition amplitudes
between states with definite numbers of
particles. Moreover, this formalism allows us to
 estimate easily  the influence of the external particles
 on the semiclassical transition.

We calculate here the probability of transition between
two coherent states.
For the purpose of calculating  multiparticle
amplitudes, the coherent state is a good approximation to the
 final multiparticle state. Unfortunately, it cannot
describe an initial two-particle state. The hope is, however,
 that the two-particle cross section can be approximated by
 the probability of transition between coherent states
 with a ``small'' number of particles in the initial state
 \cite {Approach}. Then, the problem of calculating
  the transition probability can be
 converted to the problem solving the field equations
 with some specific boundary conditions \cite {Rubakov3,Rubakov4},
 determined
 by the initial and final states. However, we cannot solve these equations
 for  arbitrary states. So  we find first any real
 solution of the equations of motion and then determine which boundary
 conditions (initial and final states)
 correspond to the solution. We will see below that these states
 and ``types'' of transition are closely related to the structure
 of the singularities of the classical solution in the complex
time plane.

First, let us consider a matrix element
\cite {Rubakov3}
$$
A_{E}(b^{\ast},a)=\langle\lbrace b_{\bf k}\rbrace
\mid SP_{E}\mid \lbrace a_
{\bf k} \rbrace \rangle
$$
which describes the amplitude for a transition at fixed energy $E$ from
the initial coherent state $\mid\lbrace a_{\bf k} \rbrace
\rangle $ (projected onto this energy
) to the final coherent state $\mid \lbrace b_
{\bf k} \rbrace \rangle $. The operator $P_{E}$
 is a projector onto subspace of
definite energy $E$; $S$ is the $S$-matrix. The system is considered
 in the center of mass frame so we do not need to project onto the
 space of definite spatial momentum.

Using the completeness condition we write this element as
$$
\langle\lbrace b_{\bf k}
\rbrace \mid SP_{E}\mid \lbrace a_
{\bf k} \rbrace \rangle=
$$
$$
=\int\,d\phi_{i}\,d\phi_{f}\,\langle
\lbrace b_{\bf k}e^{i\omega_{\bf k}T_{f}}\rbrace\mid\phi_{f}
\rangle\langle\phi_{f}\mid
U(T_{f},T_{i})\mid\phi_{i}\rangle\langle
\phi_{i}\mid P_{E}\mid\lbrace
a_{\bf k}e^{-i\omega_{\bf k}T_{i}}\rbrace \rangle
$$
In this expression $\langle
\lbrace b_{\bf k}e^{i\omega_{\bf k}T_{f}}\rbrace
\mid\phi_{f}\rangle$ and $\langle
\phi_{i}\mid P_{E}\mid\lbrace
a_{\bf k}e^{-i\omega_{\bf k}T_{i}}
\rbrace \rangle$ represent the wave functions of the
 coherent states in $\phi$ representation (the initial state is
projected onto a state of definite energy), $\phi_{i,f}
 ({\bf x})=\phi ({\bf x},T_{i,f})$ and
$T_{i}\rightarrow -\infty$,  $T_{f}\rightarrow +\infty$
are initial and final
moments of time. The matrix element of the evolution operator $U$
can be expressed by the functional integral
$$
\langle\phi_{f}\mid U(T_{f},T_{i})\mid\phi_{i}\rangle=\int\limits_{
\phi({T_{i}})=\phi_{i}}^{\phi({T_{f}})=\phi_{f}}\,D\phi\,e^{iS(\phi)}.
$$
The projection operator $P_{E}$ can be written in the form \cite{Rubakov3}
$$
\langle\lbrace b_{\bf k}
\rbrace \mid P_{E}\mid \lbrace a_
{\bf k} \rbrace \rangle=\int\,d\xi\,e^{-iE\xi}\langle\lbrace b_{\bf k}
\rbrace \mid e^{iH_{0}\xi}\mid \lbrace a_
{\bf k} \rbrace \rangle=
$$
$$
=\int\,d\xi\,e^{-iE\xi}\langle\lbrace b_{\bf k}
\rbrace \mid \lbrace a_
{\bf k}e^{i\omega_{\bf k}\xi} \rbrace \rangle=
\int\,d\xi\,\exp\lbrace -iE\xi+\int d{\bf k}b_{\bf k}^{\ast}
a_{\bf k}e^{i\omega_{\bf k}\xi}\rbrace ,
$$
where $H_{0}$ is the free Hamiltonian.

Therefore, the amplitude $A_{E}(b^{\ast},a)$, divided by the
norm of the initial state $N_{a}$
$$
N_{a}=\exp \lbrace
 {1\over 2}\int
d{\bf k}a_{\bf k}^{\ast}a_{\bf k}\rbrace
$$
 and the norm of
the final state $N_{b}$
$$
N_{b}=\exp \lbrace {1\over 2}\int
d{\bf k}b_{\bf k}^{\ast}b_{\bf k}\rbrace ,
$$
  has the following integral representation \cite {Rubakov3}
$$
A_{E}(b^{\ast},a)= \int d\xi \, d\phi_{i} \, d\phi_{f} \, D\phi
 \  exp\lbrace
-iE\xi+B_{i}(
a_{\bf k}e^{i\omega_{\bf k}\xi},\phi_{i})+
$$
\begin{equation}
\label{a}
B_{f}(b_{\bf k}^{\ast},\phi_{f})+i \int_{T_{i}}^{T_{f}}
dt{\cal L}(\phi)-{1\over 2}\int
d{\bf k}a_{\bf k}^{\ast}a_{\bf k}-{1\over 2}\int
d{\bf k}b_{\bf k}^{\ast}b_{\bf k}\rbrace .
\end{equation}

Here $\phi$ stands for all bosonic fields of the theory
 and $B_{i}(a_{\bf k},\phi_{f})$ and
 $B_{f}(b_{\bf k}^{\ast},\phi_{f})$
are the boundary terms ($\exp B_{i}$ and $\exp B_{f}$
 are the wave functions of the coherent states in the $\phi$
 representation)
$$
B_{i}(a_{\bf k},\phi_{f})= - {1\over 2} \int d{\bf k} \, a_{\bf k} \,
 a_{-{\bf k}}\,
e^{-2i\omega_{\bf {\scriptscriptstyle k}}T_{i}} - {1\over 2} \int d{\bf k}\,
\omega_{\bf
 {\scriptscriptstyle k}}\,\phi_{i}({\bf k})
\,\phi_{i}(-{\bf k})+
$$
$$
+ \int d{\bf k}\,\sqrt{2\omega_{\bf {\scriptscriptstyle k}}}\, e^{-i\omega_{\bf
 {\scriptscriptstyle k}} T{i}}\,
a_{\bf k}\,\phi_{i}({\bf k})  ,
$$
$$
B_{f}(b_{\bf k}^{\ast},\phi_{f})= - {1\over 2} \int d{\bf k}\, b^{\ast}_{\bf k}
\, b^{\ast}_{-{\bf k}}
\,e^{2i\omega_{\bf {\scriptscriptstyle k}}T_{f}} - {1\over 2} \int d{\bf k}\,
\omega_{\bf {\scriptscriptstyle k}}\,\phi_{f}({\bf k})\,\phi_{f}(-{\bf k})+
$$
\begin{equation}
\label{b}
+\int d{\bf k}\,\sqrt{2\omega_{\bf {\scriptscriptstyle k}}}\, e^{i\omega_{\bf
 {\scriptscriptstyle k}}T{f}}\,
b^{\ast}_{\bf k}\,\phi_{f}({\bf k}) ,
\end{equation}
where $\phi_{i,f}({\bf k})$ are the spatial Fourier components of the
field $\phi$ at the initial and final moments of time.

When $E \sim 1/\lambda$
and $a_{\bf k}, b_{\bf k} \sim 1/\sqrt{\lambda}$ for small $\lambda$,
we can evaluate  the transition amplitude (1) in the saddle-point
 approximation.

 The saddle-point field configuration is determined by the field equation
${\delta S / \delta \phi}=0$. The variation with respect to
 $\phi_{i}$ and $\phi_{f}$ gives
\begin{equation}
\label{c}
-i\,\dot {\phi}_{i}({\bf k})-\omega_{\bf k}\phi_{i}({\bf k})
+\sqrt {2\omega_{\bf k}}\,a_{\bf k}\,e^{-i\omega_{\bf k}T_{i}+i\omega_{\bf
 k}\xi}=0
\end{equation}
\begin{equation}
\label{d}
 i\,\dot {\phi}_{f}({\bf k})-\omega_{\bf k}\phi_{f}({\bf k})
+\sqrt {2\omega_{\bf k}}\,b_{\bf -k}^{\ast}\,e^{i\omega_{\bf k}T_{f}}=0
\end{equation}

We assume below that the field $\phi$ becomes free at large initial
 and final time, which means that its spatial Fourier transform
 can be written as a superposition of plane waves.
Therefore, we have for a  large positive time $t\rightarrow +\infty$
$$
\phi({\bf k},t)={1\over \sqrt{2\omega_{\bf k}}}\,(g_{\bf k}
\,e^{-i\omega_{\bf k}t}+{\overline {g}}_{\bf -k}\,e^{i\omega_{\bf k}t})
$$
Using the boundary conditions (4) we obtain immediately
$$
g_{\bf k}^{\ast}=b_{\bf k}^{\ast} ,
$$
i.e. the positive frequency part  of the field is determined by the
final state.

Consider now the integration with respect to $\xi$.
 The real part of $\xi$ corresponds to time translation so we
choose $\xi$ to be imaginary
$$
\xi\rightarrow\,i\,\xi
$$
and  also allow time to be complex.

The complex time formalism
has been used in quantum mechanical calculations of tunneling
events for a long time \cite {Miller}, and has been introduced
recently into instanton calculations in Ref.\cite {Rubakov3}. The
``naive''
argument in favour of this step is that we cannot describe
simultaneously the classically-allowed events (such as
free evolution of the initial and final states)
 and classically-forbidden events (for example tunneling)
in framework of the pure Minkowski or Euclidean time semiclassical
 calculations. For the classically allowed propagation, there
  exists a ``classical trajectory'', whereas a
 classically-forbidden event does not have a ``trajectory''
  in the real time).

 Consider now the contour in the complex time plane
shown on Fig.(1) \cite {Rubakov3,Rubakov4}.
 The part A of this contour is shifted upward
 and runs parallel to the real axis $t=t^{\prime}+iT$.
 Evolution of the system with respect to $t^{\prime}$
corresponds to initial state propagation,  while the real
 part of the contour describes final state propagation.
 We can interpret evolution along the imaginary part of the
 contour as some classically-forbidden event (for example,
 it can correspond under some conditions
 to a tunneling event \cite {Rubakov3}).

On part A
 of the contour, the field at early time $t^{\prime}\rightarrow -\infty$
has the form
\begin{equation}
\label{e}
\phi({\bf k},t^{\prime})={1\over \sqrt{2\omega_{\bf k}}}\,(f_{\bf k}
\,e^{-i\omega_{\bf k}t^{\prime}}+{\overline {f}}_{\bf -k}\,e^{i
\omega_{\bf k}t^{\prime}})
\end{equation}
(${\overline f}_{\bf k}$ is not complex conjugate to $f_{\bf k}$).
Substitution of this field into the first boundary condition (3) gives
$$
f_{\bf k}=a_{\bf k}\,e^{\omega_{\bf k}T-\omega_{\bf k}\xi} ,
$$
which determines the negative frequency part of the
field.

\begin{picture}(300,200)
\put (50,100){\vector(1,0){200}} \put (150,40){\vector(0,1){130}}
\thicklines \put (70,140){\line(1,0){80}}
\thicklines \put (150,140){\line(0,-1){40}}
\thicklines \put (150,100){\line(1,0){80}}
  \put (185,105){$C$}
\put (157,135){$T$} \put (155,165){Im~$t$}
\put (240,87){Re~$t$} \put (80,150){A} \put (140,0){Fig.1}
\end{picture}

Thus, to find the transition probability,
we have to solve the field equations
 with fixed negative frequency part of the field
at early time and positive frequency part of the field
at late time. This is an extremely difficult problem for
 arbitrary initial and final states, even in the
case of the $\phi^4$ theory. So we are forced to restrict
ourselves to a less general problem  \cite
 {Rubakov4}: we find first some real Minkowski-time
 solution and then find the corresponding initial and final
states as asymptotics of this solution. The ``inverted''
equations
\begin{equation}
\label{f}
b_{\bf k}^{\ast}=g_{\bf k}^{\ast}
\end{equation}
\begin{equation}
\label{g}
a_{\bf k}=f_{\bf k}\,e^{\omega_{\bf k}\xi-\omega_{\bf k}T}
\end{equation}
determine the initial and final states.

We have to make some remarks about the choice of the ``appropriate''
 solution.

First, we consider only real solutions
 because, as it has been shown in Ref.(\cite {Rubakov4}),
  the probability of the transition from the given initial state
 to all possible final states is saturated by a single final state
 which is real at real time. Therefore, the real
 saddle-point configuration
 corresponds to the transition from the given initial state to the most
probable final state. Then, we immediately obtain for the final
state $\overline {g}_{\bf k}=g_{\bf k}^{\ast}$.

The second condition is that this solution should
have an appropriate singularity structure
 in the complex time plane - we have to be able to choose the contour
 of Fig.(1) and avoid any singularities of the solution.

 We will show in the next section that $\phi^4$ theory possesses
 such solutions.

To find the amplitude of the transition we have to substitute the
saddle-point
field configuration and boundary values (6) and (7) into the integral (1).
Then, the amplitude $A$ (where $A=A_{E}(g^{\ast},f)$) is expressed by
\begin{equation}
\label{h}
A=i\int\,d\xi\,\exp\lbrace\,E\xi+iS+{1\over 2}\int d{\bf k}\,{\overline
 {f}}_{\bf k}f_{\bf k}-{1\over 2}\int\,d{\bf k}\,f_{\bf k}^{\ast}f_{\bf k}
\,e^{-2\omega_{\bf k}T+2\omega_{\bf k}\xi}\,\rbrace
\end{equation}
where we have neglected  the contributions of rapidly oscillating terms.
 In this expression $S$  represents the classical action of the
saddle-point configuration (the time integration is done along
the contour of Fig.(1)).

The integral with respect to $\xi$ can be done in the saddle-point
approximation. The saddle-point value  $\xi_{0}$ determines the energy as
a function of the other variables
\begin{equation}
\label{i}
E=\int\,d{\bf k}\,\omega_{\bf k} f_{\bf k}^{\ast}f_{\bf k}e^{-\omega_{\bf k}
 T_{0}}
\end{equation}
Here
$$
T_{0}=2T-2\xi_{0}
$$
is determined by the deviation of the shift of the contour $T$ from
 the saddle-point value of $\xi$. Equation (9) determines the parameter
$T_{0}$ in terms of energy.

After substitution of the saddle point value of $\xi$ into (8),
the probability of the transition in the saddle-point
approximation is determined by
$$
\sigma=\mid A\mid^{2}=\exp\lbrace\,2E\xi_{0}-2{\rm Im}\,S+
\int\,d{\bf k}\,{\rm Re}(\overline {f}_{\bf k}f_{\bf k})-
\int\,d{\bf k}\, f_{\bf k}^{\ast}f_{\bf k}e^{-\omega_{\bf k}
 T_{0}}\rbrace=
 $$
\begin{equation}
\label{j}
=\exp\lbrace\,-2{\rm Im}\,S+2E(T-{T_{0}\over 2})+
\int\,d{\bf k}\,{\rm Re}(\overline {f}_{\bf k}f_{\bf k})-
\int\,d{\bf k}\, f_{\bf k}^{\ast}f_{\bf k}e^{-\omega_{\bf k}
 T_{0}}\rbrace ,
\end{equation}
where ${\rm Im}S$  is imaginary part of the classical
 action, calculated
 along the time contour of Fig.(1).

Finally, we make some remarks about
 the choice of the contour of Fig.(1).
 Changing $T$ corresponds to the shift of part A of the contour upward or
downward. This shift, however, does not change the initial state
if no singularities of the solution have been crossed.
 This can be shown in the following way .  If we move
contour between lines ${\rm Im}\,t=T$ and ${\rm Im}\,t=T^
{\prime}$ (both lying between same two singularities) then the negative
frequency components of the field are related by ($\eta=T^{\prime}-T$)
$$
f_{\bf k}^{\prime}=f_{\bf k}\,e^{\omega_{\bf k}\eta} .
$$
Here we used Eq.(5) and analytically continued time $t^{\prime}
$ to  $t^{\prime}+i(T^{\prime}-T)$.
Then it is easy to see from Eq.(7) that
$$
a_{\bf k}^{\prime}=f_{\bf k}^{\prime}\,e^{\omega_{\bf k}\xi_{0}-
\omega_{\bf k}T^{\prime}}=f_{\bf k}\,e^{\omega_{\bf k}(T^{\prime}-T)}
e^{\omega_{\bf k}\xi_{0}-
\omega_{\bf k}T^{\prime}}=
$$
$$
=f_{\bf k}\,
e^{\omega_{\bf k}\xi_{0}-
\omega_{\bf k}T}=a_{\bf k} .
$$

 Contours with $T$ and $T^{\prime}$, separated by singularities,
 correspond to the completely different states (it will be shown
 in next sections
 for $\phi^4$ model, see also Refs.\cite {Rubakov4,TE}
 for other models).

At nonzero $T_{0}$, the  energy is expressed in a nonstandard
 way with respect to the negative frequency part of the field (but
it is expressed usually in terms of initial state --
 $E=\int d{
\bf k}\,a_{\bf k}^{\ast}a_{\bf k}$).  In this case  the average
 of arbitrary bilinear operator $O=\int\,d{\bf k}\,F({\bf k})
A_{\bf k}^{+}A_{\bf k}$ can be written in terms of Fourier components
of the field as \cite {Rubakov4}
$$
<O>=\int\,d{\bf k}\,F({\bf k})f_{\bf k}^{\ast}f_{\bf k}e^{-\omega_{\bf k}
T_{0}} ,
$$
where $T_{0}$ should be defined by Eq.(9). Hence, the number
 of the initial particles is expressed by
\begin{equation}
\label{k}
n_{initial}=\int\,d{\bf k}\,f_{\bf k}^{\ast}f_{\bf k}e^{-\omega_{\bf k}
T_{0}}
\end{equation}

The probability of the transition (10) does not depend on the
choice of $T$ and we can move the contour upward or downward
 until we reach a
 singularity of the classical solution. If it is possible
to choose $T$ to satisfy the condition $T_{0}=0$, which is
 equivalent to
$T=\xi_{0}$, then the energy and the number of particles are expressed
in a standard way. In this case the probability (10)
 corresponds to the expression derived in \cite {Rubakov4}.

The term $-2{\rm Im}\,S$ in the probability exponent (10) is
the suppression instanton-like factor,  while the other
 terms account for the presence of the initial particles.

In the next section we apply this formalism to the $\phi^4$ model.

\vspace{1cm}

{\section {The semiclassical process in $\phi^4$ theory}}

\vspace{1cm}

This section is devoted to consideration of
massless $\phi^4$ theory (we assume that the energy
 scale is much larger then the mass scale when the massless
 limit is a reasonable approximation).
  We describe here a real $O(4)$-invariant solution
of the theory and investigate the structure of the singularities
 of this solution in the complex time plane.
 It is shown below that the contour of Fig.(1) can
correspond to a classically-forbidden transition between two coherent
states and  the corresponding suppression factor in the transition
probability is calculated. To interpret this transition,
we analyze a similar ``process''
 in one-dimensional quantum
mechanics. Finally, as an illustration, we consider
 $\phi^4$ theory with negative coupling constant where
we can compare our results with previous calculations of the
instanton-like processes \cite {Tinyakov}.\\
{\bf 1.}\ \ \  The action of the model (we consider a
real scalar field), written
in conformally invariant form \cite {conform}, is
$$
S=\int d^4 x\, (-{1\over 2}\phi\,\partial_{\mu}\partial^{\mu}\phi-
{\lambda\over 4}\,{\phi}^{4}) ,
$$
where $\lambda>0$ is the small coupling constant. The corresponding
classical field equation is
\begin{equation}
\label{m}
\partial^2 \phi+\lambda\,\phi^3=0
\end{equation}

$O(4)$-invariant solutions of this equation \cite {Castell} can
be easily found using the invariance of the massless theory under the
Minkowski conformal group. This invariance
 can be made explicit by projecting
the theory onto the surface of the hypertorus \cite {Schechter}. Then,
$O(4)$-invariant solutions can be found by solving a one-dimensional
equation and they correspond to the oscillations with
 amplitude $a$ in
the one-dimensional potential $V(x)={1/2}x^2+{1/4}\lambda
x^4$.

 The $O(4)$-invariant solution can be expressed in terms of elliptic
functions
\begin{equation}
\label{n}
\phi(\vec x,t)={1\over \sqrt{\lambda}}
{{2a}\over {\sqrt {(r^2-(t-i)^2)(r^2-(t+i)^2)}}}\,
{\rm cn}(\sqrt {1+a^2}\,\zeta-\zeta_{0},{\it k}^2) ,
\end{equation}
where $r=\mid\vec {x}\mid$, ${\it k}^2={a^2/(2(1+a^2))}$
and
$$
\zeta={1\over {2i}}\, \ln({{r^2-(t-i)^2}\over {r^2-(t+i)^2}}) .
$$
Here {\bf cn} stands for the Jacobi elliptic cosine (see, for example,
 \cite {elliptic}) and {\it k} is
the modulus of this function. The arbitrary integration constants are
$a$ and $\zeta_{0}$. We choose $\zeta_{0}=K$ (where $K=\int_{0}^{\pi\over 2}
{{dx}/{\sqrt{
1-{\it k}^2\, \sin^{2}x}}}$ is the complete elliptic integral), in which
case $\phi=0$ at $t=0$. The constant $a$, as we will see below,
 is related to the
energy.

According to the approach, described in section (2),
 we are going to calculate
a transition corresponding to the saddle point configuration (13) considered
on the contour of Fig.(1) for some value of parameter $T={\rm Im}\,t$.

First, we investigate
 the analytic structure of the solution in the complex time plane.

This solution is real on the real time axis , so, as has been
shown in Ref. \cite {Rubakov4}, it corresponds to a transition
from the given initial state to the most probable final state.

The solution has essential singularities at $t={\pm} x {\pm} i$.
Hence,  we have to choose $T<1$ for the contour of Fig.(1) not to cross
the "light-cone" singularity.

In addition, there are
singularities (poles) at the ``points'' where
$$
\sqrt {1+a^2}\,\zeta -K=2mK+(2n+1)i\,K^{\prime}.
 $$
Here $K^{\prime} (k^2)=K(1-k^2)$, $m,n=0,\pm 1,\pm 2,...$.
 These ``points'' are poles
 of the elliptic cosine \cite {elliptic}. Because $\zeta$ is
 a function of radial coordinate and complex time, the solutions
 of this equation determine the singularity curves in the coordinate
 axes $r$, ${\rm Re}\,t$, ${\rm Im}\,t$.

We will consider below only the case $a<<1$, which, as will be shown
in the next section, corresponds to the case of a ``small'' number of
final-state
particles ($n_{final}<<{1/\lambda}$). In this limit
 $K\approx\,{\pi/2}$
 and only $m=-1$ and $n\geq{0}$ case
corresponds to the singularities in
the region ${\rm Im}\,t\geq 0$, ${\rm Re}\,t\leq 0$.
 The singularities curves (numerated by
integer number $n$) $t=t_{n}(r)$
run asymptotically ``parallel'' to the ``light-cone''
 and have $({\rm Im}\,t)$ coordinate close
 to 1
\begin{equation}
\label{o}
t_{n}=i\,\bigl(1-{({a^2\over 16})}^{2n+1}\bigr)-r
\end{equation}
at $r\rightarrow+\infty$ and $n=0,1,2...$. We have shown in Fig. (2)
  two curves in the region
 ${\rm Im}\,t\geq 0$, ${\rm Re}\,t\leq 0$.

We can see that the structure of the singularities
of this solution is ``appropriate'' --
we are able to choose the contour of Fig.(1) and not to cross
 any singularities. We choose the contour with
exactly one singularity curve under it (i.e. with $1-{a^2/16}<
T<1-({a^2/16})^3$). It will be shown below that this choice
corresponds to a classically-forbidden (exponentially suppressed)
transition.

\begin{picture}(300,300)
\put (200,100){\vector(1,0){100}}
\put (200,100){\vector(0,1){195}}
\put (200,100){\vector(-2,-1){140}}
\put (150,180){\line(-3,-1){90}}
\put (150,210){\line(-3,-1){90}}
\put (150,190){\oval(20,20)[br]}
\put (150,220){\oval(20,20)[br]}
\put (153,211){\line(-3,-1){10}}
\put (153,181){\line(-3,-1){10}}
\put (200,275){\line(-3,-1){140}}
\put (200,275){\line(-1,0){30}}
\put (150,275){\line(-1,0){30}}
\put (100,275){\line(-1,0){30}}
\put (160,190){\line(-1,0){20}}
\put (130,190){\line(-1,0){20}}
\put (100,190){\line(-1,0){20}}
\put (160,220){\line(-1,0){20}}
\put (130,220){\line(-1,0){20}}
\put (100,220){\line(-1,0){20}}
\put (90,213){\circle*{2}}
\put (90,223){\circle*{2}}
\put (90,233){\circle*{2}}
\put (70,197){$t_{1}(r)$}
\put (70,167){$t_{0}(r)$}
\put (80,30){$r$}
\put (290,87){Re~$t$}
\put (205,285){Im~$t$}
\put (193,20){Fig.~2}
\put (70,260){"$light-cone$"}
\end{picture}

The leading suppression factor in the
transition probability, according to Eq.(10), is proportional to
$$
\sigma\sim \exp(-2\,{\rm Im}\,S) .
$$
Here ${\rm Im}\,S$ should be calculated along the contour of Fig.(1).
 To calculate the imaginary part of the action we use the method of
Ref.\cite {Rubakov4}.

The action of the model is
$$
S={\lambda\over 4}\int\,d^3x\,\int\limits_{C}\,dt\,\phi^4(\vec {x},t) ,
$$
where we have used the equation of motion.  For every ${\bf x}$ the time
 integral along the contour of Fig. (1) is equal to
 to the sum of the
integral along the real time axis (which is real) and contribution of
the pole $t_{0}$, corresponding to the
singularity (14) at $n=0$. The pole contribution can be calculated
using the expression for the {\bf cn} near the singularity $-2K+iK^{\prime}$
 \cite {elliptic}
$$
{\rm cn}\,(-2K+i\,K^{\prime}+u)=-{1\over iku}-{1\over 6ik}\,(1-2k^2)\,u+O(u^2)
$$
and expanding $\zeta$ in Taylor series up to the fourth order. Poles
of the first, second and third orders give contribution to the imaginary
part of the action and after lengthy calculations the pole
contribution is equal to the integral
\begin{equation}
\label{p}
{\rm Im}\,S={\rm Im}\,S_{pole}={2\pi^{2}\over \lambda}\
\int\limits_{0}^{\infty}\
{-80\,t^3\,(r^2-(t-i)^2)^{2}(r^2-(t+i)^2)^{2}\over {(1+t^2+r^2)^{7}}}
\, r^2\,dr
\end{equation}
evaluated at $t=t_{0}(r)$. After substitution of the exact equation
of the singularity line in the form
$$
t_{0}=p-\sqrt {p^2+r^2+1}
$$
(where $p$ is $i\,(1-{a^2/16})$ for small $a$) the integral (15) is
reduced to the integral
\begin{equation}
\label{q}
{\rm Im}\,S={20\pi^2\over \lambda}\,\,\int\limits_{0}^{\infty}
\,{(1+p^2)^2\,r^2\over {(1+r^2+p^2)^{7/2}}}\, dr
\end{equation}

By substitution $y^2={r^2/{(1+p^2)}}$ factor $p$ can be scaled out
of integral and the final result is
\begin{equation}
\label{r}
{\rm Im}\,S= {20\pi^2\over \lambda}\, \,\int\limits_{0}^{\infty}
\,{y^2\,dy\over (1+y^2)^{7/2}}={8\,\pi^2\over {3\,\lambda}}
\end{equation}

It is exactly equal to Lipatov instanton action: the
 Euclidean action of the classical solution in $\phi^4$ theory with
 negative coupling constant \cite {Lipatov} (our normalization of
$\lambda$ differs from the normalization of $\lambda$ in \cite {Lipatov}
by factor 6).
 Thus, the choice of the contour between the first and the
second singularity line corresponds to the classically
forbidden transition suppressed by the factor
$$
\sigma\,\sim\,\exp(-2\,S_{0}) ,
$$
where $S_{0}$ is equal to Lipatov instanton action.
 So this process is analogous to the one-instanton transition in
the electroweak model or to the ``instanton-like'' transition in $\phi^4$
 theory with negative coupling constant.

The existence of such ``instanton-like'' processes in the $\phi^4$ theory
with  positive coupling constant seems surprising. However,
 we can find analogy in one-dimensional
quantum mechanics.\\
{\bf 2.}\ \ \ Let us consider scattering above
the potential barrier $V(x)$ in
the semiclassical approximation, following the approach described
in the textbook of Landau and Lifshitz \cite {Landau}.
 The transmission above the
 barrier is classically-allowed, so, in the first approximation,
 the transmission probability is equal to one $T=1$ and the reflection
probability is zero (i.e. exponentially small) $R=0$. We know,
however, that there should exist some classically forbidden reflection from
the barrier.

According to the general approach of \cite {Landau}, to calculate the
classically-forbidden probability of transition from some initial
state to some final state we have to find classical ``trajectory''
connecting initial and final ``points''  and calculate action $S(q_{1},
q_{0})+S(q_{0},q_{2})$ (the classical action is $S=\int\,p\,dx$,
  where $p(x)=\sqrt{2m(E-V(x))}$ is a classical momentum)
for the evolution of the system from the initial
``point'' $q_{1}$ to the ``turning point'' $q_{0}$ (singular
 point of the classical momentum) and then from the $q_{0}$
to the final ``point'' $q_{2}$. Then, the probability of the process is
proportional to
$$
\omega\,\sim\,\exp\lbrace-{2\over \hbar}\,{\rm Im}(S_{1}(q_{1},
q_{0})+S_{2}(q_{0},q_{2})\rbrace .
$$

In our case the singular ``turning point'' is some
complex coordinate $x_{0}$ (and complex conjugate
coordinate $x_{0}^{\ast}$)
 determined by the requirement $V(x_{0})=E$. The  classically-allowed
contribution to the transmission probability corresponds to
the action on the trajectory
along the real $x$-axis (or a trajectory which can be deformed
to the real axis) and
connects points $x_{1}\rightarrow -\infty$ and $x_{2}\rightarrow +\infty$
(Fig.(3), trajectory A).

This
trajectory gives a contribution equal to
 one to the transmission probability.

The classically forbidden reflection is  determined by the trajectory
in the complex coordinate plane
which connects points $x_{1},x_{2}\rightarrow -\infty$ and ``winds'' around
the  ``turning point'' $x_{0}$ (Fig.(3), trajectory B). The reflection
probability is determined by the imaginary part of the classical action on
this trajectory
$$
R=\mid A_{R}\mid^2\,\sim\,\exp\,(-{2\over \hbar}\,{\rm Im}
\int\limits_{C}p\,dx) ,
$$
where $A_{R}$ is the amplitude of reflection.

\begin{picture}(300,300)
\put (50,150){\vector(1,0){200}}
\put (150,50){\vector(0,1){230}}
\put (75,150){\circle*{3}}
\thicklines \put (75,150){\vector(1,0){75}}
\thicklines \put (150,150){\line(1,0){75}}
\put (225,150){\circle*{3}}
\put (87,155){\circle*{3}}
\thicklines \put (87,155){\vector(1,0){63}}
\thicklines \put (150,155){\line(1,0){70}}
\thicklines \put (150,155){\oval(140,140)[t]}
\put (80,155){\circle*{3}} \put (70,160){\circle*{3}}
\thicklines \put (70,160){\vector(1,0){80}}
\thicklines \put (150,160){\line(1,0){85}}
\thicklines \put (150,160){\oval(170,170)[t]}
\thicklines \put (150,140){\oval(170,170)[b]}
\thicklines \put (65,160){\line(0,-1){20}}
\put (235,140){\circle*{3}} \put (250,137){Re~$x$}
\put (150,105){\circle*{5}}
\put (160,105){$x_{0}^{\ast}$}
\put (150,195){\circle*{5}} \put (160,195){$x_{0}$}
\put (157,260){Im~$x$} \put (140,10){Fig.~3}
\put (120,140){A} \put (120,215){B} \put (120,235){C}
\end{picture}

Because for the $T=1$ and $R\not=0$ the unitarity condition $R+T=1$
is violated, in order to ``unitarize'' the amplitudes
we have to take into account the classically forbidden
contribution to the transmission amplitude corresponding
to the ``transmission
 after reflections''. This contribution is determined by the trajectory
connecting points $x_{1}\rightarrow -\infty$ and
 $x_{2}\rightarrow +\infty$ and
``winding'' around the ``turning points''
 $x_{0}$ and $x_{0}^{\ast}$. This
contribution to the amplitude of the transmission is proportional to
$$
A_{T}\sim\,\exp(-{1\over \hbar}\,{\rm Im}\,\int\limits_{C}p\,dx) ,
$$
where $C$ is a contour C on the Fig.(3).

Thus, the transmission amplitude  is dominated in the semiclassical
approximation by the saddle-point contributions,
corresponding to the trajectories in the complex coordinate plane.
The trajectory along the real axis (or a
trajectory which can be deformed to the
real axis) corresponds
 to the classically allowed contribution, and trajectories, which cannot
be deformed to the real axis (without crossing the singular
``turning points''), describe classically-forbidden contributions
corresponding to ``transmissions after reflections''.

Of course, we cannot relate directly quantum mechanical
 and $\phi^4$ field theory examples. But both models
 have a common feature, namely that the
 amplitude of transition in the semiclassical approximation
 is dominated by complex saddle-point configurations, representing
 a classical solution analytically continued to the complex ``coordinate''
plane. These contributions can be ``classified'' by the position of the
 corresponding ``trajectories'' with respect to the singularities
 of this solution.

Part A of the contour of Fig.(1) describes the free propagation
 of an incoming spherically-symmetric shell (13) at early time.
 Evolution along the imaginary part of the contour can be interpreted
 as a classically-forbidden reflection in the $\phi^4$ potential. The
 Minkowski part of the contour corresponds to an outgoing wave
 at late time. Therefore,
  we can call the
semiclassical process in the $\phi^4$ model, described by the
nontrivial ``trajectory'' (lying between the singular lines)
in the complex time plane, as a
``transmission after reflections''. Like the quantum mechanical
example, including the contribution of such processes into the
 corresponding amplitude can, probably, unitarize the perturbative
amplitude.\\
{\bf 3.}\ \ \ Now, as an illustration,
we want to investigate the $\phi^4$ model
with negative coupling constant. This theory allows instanton-like
processes, which can be considered as  models for the ``shadow processes''
 \cite {Voloshin-Hsu} (processes describing transitions from initial
particles in the false vacuum to final-state particles in the false vacuum
 through an intermediate state containing a bubble of the true vacuum).
The probability of such processes has been derived in the framework
 of the instanton formalism in Ref.\cite {Tinyakov}. It is proportional
to the $\sigma\,\sim\,\exp(-2S_{0})$, where $S_{0}$ is the Lipatov
instanton action. We will show below that such processes can be
 described by the classical solution considered in the complex time plane
(at least up to the suppression factor in the
transition probability).

We analyze here only the massless case. Of course, a mass term has to be
added to make the $\phi=0$ state at least metastable,  but we expect
that the mass corrections
 to the transition probability shall not  affect the
 leading suppression term
 $\sigma\sim \exp(-2\,{\rm Im}\,S)$.

The $O(4)$-invariant classical solutions of Eq.(2) with
negative coupling constant $\lambda=-\mid\lambda\mid$ can be found
 using the approach of Ref.\cite {Schechter}. After ``reduction''
 of the theory to a one-dimensional model, they correspond
 to oscillations with an amplitude $a$ in the one-dimensional
 potential $V={1/2}\phi^2-{\lambda/4}\phi^4$
(so even massless theory is metastable in the ``subspace'' of the
$O(4)$-invariant solutions). The solution
can be written in the form
$$
\phi(\vec x,t)={1\over \sqrt{\mid \lambda \mid}
}{{2a}\over {\sqrt {(r^2-(t-i)^2)(r^2-(t+i)^2)}}}\,
{\rm sn}(\sqrt {1-{a^2\over 2}}\,\zeta-\zeta_{0},{\it k}^2) ,
$$
where again $r=\mid\vec {x}\mid$ and
 $\zeta={1/{(2i)}}\, \ln({{(r^2-(t-i)^2)}/{(r^2-(t+i)^2)}})$,
 ${\it k}^2={a^{2}/{(2-a^2)}}$. We choose $\zeta_{0}$ to be $K$
, which corresponds to the case ${\partial\phi/\partial{t}}=0$
 at $t=0$. One can see that the structure of the singularities in the
complex time plane does not change relative to the case with positive
coupling constant. Choosing the contour between the first and second
singularity lines, and using the expansion of the elliptic sin function
near the singularity
$$
{\rm sn}\,(-2K+i\,K^{\prime}+u)
=-{1\over ku}-{1\over 6k}\,(1+k^2)\,u+O(u^2) ,
$$
we obtain the result that the imaginary part of the action on this contour
is exactly equal to the Lipatov instanton
 action $S_{0}={8\pi^2/(3\mid\lambda
\mid)}$. Thus, the transition probability of this process
 is suppressed by a
factor $\sigma\,\sim\,\exp(-2S_{0})$ and this process indeed
 describes (at
 least up to the suppression factor in the probability)
 a transition in the ``one-instanton'' sector, previously considered
in Ref.\cite {Tinyakov}.

\vspace{1cm}

{\section {The initial and final states}}

\vspace{1cm}

In this section we calculate the Fourier components of the initial
 and final states and find the corresponding
energy and average number of
particles. As has been mentioned before, the final state is determined
 via Eq.(6) by the asymptotics of the classical solution on the Minkowski
part of the contour of Fig.(1) in the limit $t\rightarrow +\infty$, while
 the initial state corresponds to the asymptotics of the solution on
 part A of the contour in the limit $t^\prime\rightarrow -\infty$,
 where $t=iT+t^\prime$ (Eq.(7)).
We consider only case $a<<1$ which, we will see below, corresponds
to the case $n_{final}<<{1/\lambda}$.

First, we determine the final state. In the limit $a<<1$, the Fourier
components can be easily found by using the first-order approximation for
the elliptic cosine ($k$ here is a modulus of the elliptic
function)
\begin{equation}
\label{s}
{\rm cn}(u,k)\mid_{k\rightarrow 0}=\cos\,u
\end{equation}

The Fourier components are
$$
g_{{\bf k}}=a\,\sqrt{{2\pi\over 2\,{\rm k}\lambda}}\,i\,e^{-{\rm k}} ,
$$
where ${\rm k}=\mid {\bf k}\mid$,
for the negative frequency part of the field and
$$
\overline {g}_{{\bf k}}= g_{{\bf k}}^{\ast}=-\,
a\,\sqrt{{2\pi\over 2\,{\rm k}\lambda}}\,i\,e^{-{\rm k}}
$$
for the positive frequency part of the field.

The energy, the number of final particles $n_{final}$ and
the average momentum ${\rm k}_{average}\approx\,{E/n_{final}}$
are determined as
$$
E=\int\,d{\bf {k}}\,{\rm k}\,g_{\bf k}^{\ast}\,g_{\bf
 k}=\,{\pi^2\,a^2\over \lambda}
$$
$$
n_{final}=
\int\,d{\bf k}\,g_{\bf k}^{\ast}\,g_{\bf
  k}=\,{\pi^2\,a^2\over \lambda}
$$
and
$$
{\rm k}_{average}^{initial}\sim 1 .
$$

To find the initial state is not as easy.
The initial state should be determined by
the asymptotics of the solution on
part A of the contour of Fig.(1). In this case we cannot use a first-order
approximation (18) since we have to integrate through the region in the
vicinity of the singularity of the elliptic cosine. Instead, we consider
the integral determining the Fourier components of the solution
\begin{equation}
\label{t}
\phi({\bf k},t)={\sqrt{2{\rm k}}\over (2\pi)^{3/2}}\,\int\,d^{3
}{\bf  x} \phi({\bf  x},t)\,e^{i{\bf  k  x}}=
\sqrt{{2\pi\over 2\,{\rm k}}}\,\int\limits_{-\infty}^{+\infty}
\phi({\bf x},t)\,e^{i{\rm k}r}{r\over i{\rm k}}\,dr
\end{equation}
in the complex plane of the variable $r$. The solution has complicated
singularity structure at complex $r$ , including poles and branch points.
 However, in the case $a<<1$ , we can calculate leading contributions
to the Fourier components using the following trick.

Consider again the final state. If $\phi({\bf  x},t)$ is
 an exact solution, the integral (19),
 calculated along the real $r$-axis,
corresponds to  the exact Fourier components of the solution. The solution
$\phi({\bf  x},t)$ has poles at points $r=\pm i\,(1-({a^2/16})^{
2n+1})\pm\,t$ and branch points at $r=\pm\,i \pm\,t$. If we want to calculate
 the Fourier components of the initial state we have to make an analytical
continuation of the integral (19) to the complex time $t\rightarrow\,t+iT$.
 As the result of this continuation the
pole $r=-t+i(1-{a^2/16})$ crosses the real $r$-axis from above
and the pole $r=t-i(1-{a^2/16})$ crosses the real $r$-axis
 from below
(other singularities do not cross the real $r$-axis). For $T\equiv 1-
\epsilon$ the expression for the ``new'' position of the poles is
$$
r=-t+i(1-{a^2\over 16})\longrightarrow\,r({\rm I})=
-t-i({a^2\over 16}-\epsilon)
$$
\begin{equation}
\label{u}
r=t-i({a^2\over 16})\longrightarrow\,r({\rm II})=
t+i({a^2\over 16}-\epsilon) .
\end{equation}

\begin{picture}(300,200)
\put (50,100){\vector(1,0){200}}
\put (150,40){\vector(0,1){130}}
\thicklines \put (50,105){\line(1,0){20}}
\put (85,105){\circle*{5}}
\thicklines \put (85,105){\oval(30,30)[t]}
\thicklines \put (100,105){\vector(1,0){50}}
\thicklines \put (150,105){\line(1,0){50}}
\thicklines \put (215,105){\oval(30,30)[b]}
\put (215,95){\circle*{5}}
\thicklines \put (230,105){\line(1,0){15}}
\thicklines \put (50,100){\vector(1,0){100}}
\thicklines \put (150,100){\line(1,0){45}}
\put (85,105){\vector(0,-1){13}}
\put (215,95){\vector(0,1){13}}
\put (245,87){Re~$r$} \put (157,160){Im~$r$}
\put (140,0){Fig.~4} \put (105,115){I}
\put (230,90){II} \put (130,110){$initial$}
\put (132,90){$final$}
\end{picture}

We can see that the difference between the
Fourier components of the final state,
 analytically continued to complex time, and the Fourier components
of the initial state is given by the contributions of the poles (18)
 to the integral (19) (see Fig.\,(4)).

 We can write it  symbolically  as
\begin{equation}
\label{v}
\phi_{initial}({\bf k},t)=\phi_{final}({\bf k},t\rightarrow
\,t+iT)\,-\,pole({\rm I})\,+\,pole({\rm II}) ,
\end{equation}
where $pole({\rm I,\,II})$ presents the contributions of
 poles to the integral (19) and $\phi_{final}$ is the contribution of the
final state, analytically continued to complex time.
 In the leading approximation at small $a$
 we can use again the approximation (18) to calculate the contribution
of the final state in Eq.(21).

 To calculate contributions of the poles,
we find the residues of the classical solutions at the singular points
$$
res({\rm I}),res({\rm II})=-{\sqrt{2}\,i\over a}\,{\sqrt {(r-t-i)
(r-t+i)(r+t+i)(r+t-i)}\over (-4\,t\,r)}
$$
and substitute $r=r({\rm I}),r({\rm II}); t\rightarrow\,t+iT$.
Here we use the expression for the residue of the elliptic cosine \cite
{elliptic},  which equals $-{i/k}$, and corresponds to the
singularities determined by
 numbers $m=0,\,n=0$ in Eq.(14).

The contribution of $pole({\rm I})$ cancels the leading
term, proportional $a$, in the first term of Eq.(21).
 We obtain the leading contribution to the
negative frequency part of the field
 in the limit ${\rm Re}\,t\rightarrow\,-\infty$,
$$
f_{\bf k}\approx \sqrt{{2\pi\over 2{\rm k}\lambda}}\,i\,(a\,e^{-{\rm k}
\epsilon}-a\,e^{-{\rm k}\epsilon+{\rm k}a^2/16})\approx
$$
$$
\approx -\sqrt{{2\pi\over 2{\rm k}\lambda}}\,i\,{a^3\over 16}{\rm k}\,e^{-{\rm
k}
\epsilon} .
$$
In this expression we neglect terms of the order $a^3$ in the final state
 contribution, coming from the $a^2$ term in the expansion of the elliptic
 cosine, because these terms are proportional to
${1/\sqrt {{\rm k}}}e^{-
{\rm k}\epsilon}$ and give suppressed contributions to the energy and number
 of particles.

The positive frequency part of the initial state is proportional to $a$
$$
{\overline {f}}_{\rm k}\approx \sqrt{{2\pi\over 2{\rm k}\lambda}}
i\,a\,(-e^{-(2-\epsilon){\rm k}}+e^{-{\rm k}a^2/16+\epsilon{\rm k}}) .
$$

To find number of the initial particles we use Eq.(11) because
 we cannot choose T (or $\epsilon$) to satisfy condition $E=\int\,d{\bf k}
{\rm k}\,f_{\bf k}^{\ast}f_{\bf k}$. The parameter $T_{0}$ in Eq.(11)
 is determined by the requirement
\begin{equation}
\label{w}
E=\int\,d{\bf k}\,
{\rm k}\,f_{\bf k}^{\ast}f_{\bf k}\,e^{-{\rm k}T_{0}}
\end{equation}
After substitution of the negative frequency part of the Fourier components
 into (22), we obtain
$$
E={\pi^2\,a^2\over \lambda}={3\,\pi^2 a^6\over 4^4\,(\epsilon+T_{0}/2)^5
\,\lambda} .
 $$
This gives us the expression for $T_{0}$,
$$
\epsilon+{1\over 2}\,T_{0}=({3\over {4^4}})^{1/5}\,a^{4/5}
$$
The number of the initial particles is given by (11) and equals
$$
n_{initial}=\int\,d{\bf k}
\,f_{\bf k}^{\ast}f_{\bf k}\,e^{-{\rm k}T_{0}}
$$
$$
={1\over 2}\,E\,(\epsilon+{1\over 2}\,T_{0})=({3\over 4^4})^{1/5}{\pi^2\,
a^2\over 2\,\lambda}
\,a^{4/5}={1\over 2}\,({3\over 4^4})^{1/5}\,a^{4/5}\,n_{final} .
$$
This result implies
$$
n_{final}\sim\,{n_{initial}^{5/7}\over \lambda^{2/7}}
$$
and
$$
{\rm k}_{average}^{initial}\sim\,{{\rm k}_{average}^{final}\over a^{4/5}}.
$$

We can see that for small coupling constant the number
of the final ``soft'' particles is much larger then the number of
the initial ``hard'' particles.

Thus, the classical solution, considered on the contour of Fig.(1) in
the complex time plane above the singularity line, corresponds to the
 transition between two coherent states with a ``strong''
 violation of particle number, $n_{final}>>n_{initial}$.

\vspace{1cm}
{\section {Concluding remarks}}
\vspace{1cm}

In the previous sections we have studied the
semiclassical process in $\phi^4$ theory with positive
coupling constant, which describes transition
between two coherent states. This transition is suppressed
 by the factor
$\exp (-2S_{0})$, where $S_{0}$ is equal to the Lipatov
instanton action -- the Euclidean action of the classical solution
 in the  theory with negative coupling constant.

The initial and final states, corresponding to this transition
, have different numbers of particles ($n_{final}>>n_{initial}$)
 and different average momenta (${\rm k}_{final}<< {\rm k}_{initial}$),
 so this transition approximates some multiparticle
scattering process with a large number
 of ``soft'' final particles.

 The process is technically analogous to the one-instanton
 transition
in electroweak model and could serve as a
good model for studying the instanton effects. It seems that we can
also describe some ``multi-instanton'' processes using the solution
 (13) and choosing the contour of Fig.(1) above several singularity
lines.

We believe that we have to include the contributions
 of these instanton-like
processes into the corresponding ``total'' amplitude for multiparticle
production. Such contributions might slow down the
 factorial growth of the perturbative amplitude
and unitarize the high energy cross section.

The energy dependence of the transition probability
 of this process is a very interesting problem. It requires a detail
investigation of Eq.(10).  The growth of the transition probability
is related to the presence of the external
particles. The first term in Eq.(10) describes the suppression
 factor while other terms describe the contributions
of the external particles. These terms are
 trying to overcome the suppression factor and can be, in
 principle,  large.  An accurate estimation of the contribution
 of the external particles requires, however,
  including mass term effects
 into consideration.

We have to add the mass term for the following
  reason.
  Calculation of the
 transition probability requires summing the
contributions from different ``sizes'' of the
classical field. In this paper we consider only the contribution
 of the solution with
 a ``unit'' size (field configuration (13)).
 This integration is divergent at the large
 ``sizes'' and should be regularized by introducing
  a mass term into the action in the manner of the ``constrained
 instanton'' approach \cite {Affleck}.
We do not consider the effects of the mass term in this paper, so
 this problem requires a more detailed investigation.

The important point is that the
 framework of the formalism allows one to analyze, in principle,
  the case $n_{final}\ge {1/\lambda}$. This case is analogous
 to multi-particle scattering at the sphaleron energy
in the standard model, where the behavior of multi-particle
 cross section is still far from being understood.

\vspace{1cm}
{\section*{Acknowledgments}}

 The author is indebted to
 T. Gould and E. Poppitz for stimulating
discussions and critical remarks at all stages of this work, especially
regarding the problem of calculation of the
Fourier components of the initial state.
 I am grateful to J. Bagger and G. Domokos for
valuable discussions and interesting suggestions.

\end{document}